\documentclass[review]{elsarticle}
\usepackage{lineno}
\usepackage{xcolor}
\usepackage[margin=1in]{geometry}
\usepackage{hyperref}
\hypersetup{hidelinks}
\usepackage{algorithm}
\usepackage{algpseudocode}

\usepackage{amsmath}
\newcommand{\bfm}[1]             { \mathbf{#1}     }       
\newcommand{\uu}              { \bfm{u} } 
 
\usepackage{booktabs}       % professional-quality tables
\usepackage{longtable}
\usepackage{amssymb}

\journal{Coastal Engineering}

\begin{document}

\begin{frontmatter}

\title{A Framework for Flexible Peak Storm Surge Prediction}

%\date{September 9, 1985}	% Here you can change the date presented in the paper title
%\date{} 					% Or removing it
%Pachev \href{https://orcid.org/0000-0002-9067-0491}{\includegraphics[scale=0.06]{orcid.pdf}\hspace{1mm}
%Dawson \href{https://orcid.org/0000-0001-7273-0684}
%Valseth \href{https://orcid.org/0000-0001-6940-4191}
\author[inst1]{Benjamin Pachev\corref{cor1}}
\ead{benjaminpachev@utexas.edu} 
\affiliation[inst1]{
  institution={Oden Institute for Computational Engineering and Sciences, The University of Texas at Austin},
  city={Austin},
  postcode={78712},
  state={Texas},
  country={USA},
}

\author[inst5]{Prateek Arora}
\ead{pa2178@nyu.edu}
\affiliation[inst5]{
  institution={Department of Civil and Urban Engineering, New York University,},
  city={Brooklyn},
  postcode={11201},
  state={New York},
  country={USA},
}

\author[inst1]{Carlos del-Castillo-Negrete}
\ead{carlos.delcastillo@utexas.edu}

\author[inst1,inst3,inst4]{Eirik Valseth}
\ead{eirik.valseth@nmbu.no,eirik@oden.utexas.edu} 
\affiliation[inst3]{organization={The Department of Data Science, Norwegian University of Life Sciences},%Department and Organization
            addressline={Elizabeth Stephansens vei 15}, 
            city={Ås},
            postcode={1430}, 
            country={Norway}}
\affiliation[inst4]{organization={Simula Research Laboratory},%Department and Organization
            addressline={Kristian Augusts gate 23}, 
            city={Oslo},
            postcode={0164}, 
            country={Norway}}

\author[inst1]{Clint Dawson}
\ead{clint@oden.utexas.edu}

 \cortext[cor1]{Corresponding author}

% Uncomment to remove the date
%\date{}

% Uncomment to override  the `A preprint' in the header
%\renewcommand{\headeright}{Technical Report}
%\renewcommand{\undertitle}{Technical Report}
%\renewcommand{\shorttitle}{\textit{arXiv} Template}

%%% Add PDF metadata to help others organize their library
%%% Once the PDF is generated, you can check the metadata with
%%% $ pdfinfo template.pdf
%\hypersetup{
%pdftitle={Learning Storm Surge with XGBoost},
%pdfsubject={eess.SP, cs.LG, stat.ML},
%pdfauthor={Benjamin Pachev, Clint Dawson},
%pdfkeywords={Storm Surge, ADCIRC, Machine Learning},
%}

\begin{abstract}

Storm surge is a major natural hazard in coastal regions, responsible both for significant property damage and loss of life. Accurate, efficient models of storm surge are needed both to assess long-term risk and to guide emergency management decisions. While high-fidelity regional- and global-ocean circulation models such as the ADvanced CIRCulation (ADCIRC) model can accurately predict storm surge, they are very computationally expensive. Consequently, there have been a number of efforts in recent years to develop data-driven surrogate models for storm surge. 

Here we develop a novel surrogate model for peak storm surge prediction based on a multi-stage approach. In the first stage, points are classified as inundated or not. In the second, the level of inundation is predicted for each point. Additionally, we propose a new formulation of the surrogate problem in which storm surge is predicted independently for each point. This formulation allows for predictions to be made directly for locations not present in the training data, and significantly reduces the number of required model parameters.
%We find the two-stage approach yields better accuracy than a one-stage model. Furthermore, it allows direct prediction of the zone of inundation in addition to determining the maximum water elevation.

We demonstrate our modeling framework on two study areas: the Texas coast and the northern portion of the Alaskan coast. For Texas, the model is trained with a database of 446 synthetic hurricanes. The model is able to accurately match ADCIRC predictions on a test set of synthetic storms. We further present a test of the model on Hurricanes Ike (2008) and Harvey (2017). For both storms, we find that the model predictions have comparable accuracy to ADCIRC hindcasts when compared to actual observational data. For Alaska, the model is trained on a dataset of 109 historical surge events. We test the surrogate model on actual surge events including the recent Typhoon Merbok (2022) that take place after the events in the training data. As with the Texas dataset, the surrogate model achieves similar performance to ADCIRC on real events when compared to observational data. In both cases, the surrogate models are many orders of magnitude faster than ADCIRC.

\end{abstract}


%\begin{highlights}
%%%%%%%%%%%%%%%%%%%%%%%%%%%%%%%%%%%%%%%%%%%%%%%%%%%%%%%%%%%%%%%%%%%%%%%%%%%%%%%%%%%%%%%%%%% Last % is 85 Characters starting one space after \item
%\item We introduce a new machine-learning model for predicting storm surge in Texas  % with a mean accuracy of .25 meters.
%\item The mean accuracy of the novel machine-learning surrogate is 0.25 meters
%\item Validation shows that training with synthetic storms is insufficient for real storms
%\item We demonstrate that synthetic storms typically used for training surrogate models fail to capture the full range of real hurricanes.
%\item Using perturbations of real storms in training significantly improves the performance % for actual hurricanes over the purely synthetic baseline.
%\end{highlights}

\end{frontmatter}
%\linenumbers

\section{Introduction}
In the last four decades, tropical cyclones have caused over one trillion dollars of damage in the United States alone~\cite{noaabillion}. Storm surge is directly responsible for much of the property damage from tropical cyclones~\cite{neumann2015joint} and nearly half of the fatalities~\cite{rappaport2014fatalities}. Hurricane Katrina (2005) was the costliest hurricane on US record, with hundreds of billions of dollars in property damage and over 1200 deaths  most of which were caused by the extreme storm surge~\cite{blake2011deadliest}. Thus, predicting storm surges is crucial in order to assess the long term risk to coastal infrastructure and property from tropical cyclones. While high-fidelity physics-based models of storm surge such as the ADvanced CIRCulation (ADCIRC) model~\cite{luettich1992adcirc,pringle2021global} have been developed, they require significant computational resources. This makes statistical risk studies infeasible or limited in scope. One workaround is to use a fast, low-fidelity physical model such as SLOSH (Sea, Lake, and Overland Surges from Hurricanes,~\cite{jelesnianski1992slosh}). However, low-fidelity models neglect key physics and consequently can have high errors.

Researchers have focused on deploying Machine Learning (ML) and Deep Learning (DL) based storm surge prediction models, e.g.,~\cite{LEE2006483,LEE200863,LEE20091200, ian2022performance, ALKAJBAF2020106184, nhess-12-3799-2012, kim2015time, hashemi2016efficient, DBLP:journals/corr/BezuglovBS16, kim2018surrogate, LEE2021104024}, which aim to mimic the physics of storm surge as accurate high fidelity models, but with a significant reduction in computational requirements. In~\cite {ian2022performance}, Ian \emph{et al.} compared the performance of several machine learning models to predict storm surge and observed that tree-based ensemble methods provided the best estimates of storm surge. Whereas in~\citep{ALKAJBAF2020106184}, Al Kajbaf and Bensi compared the performance of DL-based Artificial Neural Networks (ANN) and ML-based Gaussian Process Regression (GPR) and Support Vector Machines (SVM) methods to predict storm surge levels. Finally, Al Kajbaf and Bensi  predicted the storm surge with the least errors using ANN compared to GPR and SVM methods in~\citep{ALKAJBAF2020106184}. 
% \eirik{least compared to what?}
Lee \emph{et al.}  used ANN, including Back Propagation Neural Network (BPNN), to model the storm surge from Typhoons in Taiwan in \citep{LEE2006483, LEE200863, LEE20091200} using the inputs for storm characteristics which included: wind velocity, wind direction, and atmospheric pressure. Furthermore, these studies and also include harmonic analysis of tidal levels. However, these data-driven storm surge models  had limited training data and could not be generalized or valid during different tidal seasons. In~\citep{nhess-12-3799-2012}, Chen \emph{et al.} explored the ANN structures, which included BPNN and Adaptive Neuro-Fuzzy Inference Systems (ANFIS) algorithms to predict storm surge at a few tidal gauges on the east coast of Taiwan based on four input parameters: wind speed, wind direction, air pressure and simulated water level from ADCIRC.  The ANFIS model outperformed BPNN in all cases.

More advanced ML techniques were developed by Kim \emph{et al.} in~\citep{kim2015time} in which a larger set of six input parameters were used:  latitude, longitude, the central pressure of the storm, moving speed of the storm, heading direction of the storm, and radius of exponential scale pressure. These parameters were used in a feed-forward neural network for a dataset of 446 synthetic storms developed using historical tropical cyclone activity in the Gulf of Mexico to develop a surrogate storm surge prediction model. In \citep{hashemi2016efficient}, Hashemi \emph{et al.} used 1050 synthetic storms simulated by US Army Corp of Engineers for the North Atlantic Comprehensive Coastal Study to predict storm surge and compared the performance of DL based ANN, and ML based Support Vector Machine (SVM) algorithms. The developed prediction model was applied to weather stations within Rhode Island, USA, using four storm parameters: central pressure deficit, the radius of maximum winds, translation speed, and storm heading direction. The results reported in \cite{hashemi2016efficient} indicate that ANN performed better than SVM for storm surge predictions in the considered scenarios. 
%Bezuglov \emph{et al}  included the time to landfall as an additional parameter in~\citep{DBLP:journals/corr/BezuglovBS16} to develop a surrogate ANN based model for storm surge predictions in North Carolina, USA. \eirik{how did this additional parameter impact the predictions?} 
%\eirik{these two following references should be better explained. Now it looks like we are just mentioning them for the sake of it. Problably we can remove them altogether as the lit. review is very thorough} \citep{kim2018surrogate} developed the ANN based storm surge model for South Korea. \citep{LEE2021104024} employed data for 1031 synthetic storms to predict storm surge predictions in Chesapeake Bay.

Overall, DL and ML based surrogate models have been capable of predicting storm surge with reasonable accuracy. However, the currently existing models, e.g. \cite{nhess-12-3799-2012, kim2015time, hashemi2016efficient, kim2018surrogate, LEE2021104024} ignore the complex bathymetry and topography of coastal regions which can introduce uncertainties in storm surge prediction, e.g., in Alaska, USA where  storm-tidal interactions are significant \citep{zhang2021effects}. To overcome these shortcomings, we developed a high-fidelity surrogate storm surge prediction model which is not limited to storm characteristics alone, and includes complex bathymetry and topography features of coastal regions.
Improved ML driven predictions of storm surge can be used to determine the expected performance of existing and planned infrastructure as well as during preparations for incoming hazards \citep{egusphere-2022-975}. 
% \textcolor{red}{
% Need to bring some description about dataset and process here.
% }
% \textcolor{blue}{The dataset is extensively discussed later in the paper . . .}

In this paper, we present a novel flexible framework to develop  regional surrogate models for peak storm surge prediction.
The framework is applied to the Texas and northern Alaskan coastlines. The surrogate models rely on Gradient Boosting ML and ANN-based DL algorithms.  A total of  135 and 172 input features are used for Texas and Alaska, respectively. The input features include bathymetric features, tidal harmonics, wind, pressure, as well as other storm characteristics. The trained models are validated on the historic events of Hurricane Harvey (2017), and Hurricane Ike (2008) in Texas and Typhoon Merbok (2022) in Alaska. The remainder of the paper is organized as follows: in Section \ref{sec:datasets} we discuss the study areas and creation of corresponding datasets. Section~\ref{sec:model}, we detail the construction of surrogate models. In Section~\ref{sec:results}  we present our numerical results for both study areas. Finally, we conclude with a discussion of our results and future work in Section~\ref{sec:conclusion}.

\section{Dataset Generation}
\label{sec:datasets}

There are two potential sources of training data for a machine learning model of storm surge: observations, or output from a high resolution numerical model. Observational data is limited to only a few points in the spatial domain (e.g NOAA or USGS tide gauges). Consequently, it's not feasible to train a model with full spatial output from observational data alone. This necessitates the use of a physics-based model to generate training data with significantly larger spatial extent. However, it is worth noting that a data-driven model, while much computationally cheaper to evaluate, will inherit the same biases and errors present in the training data. Consequently, the model used for data generation should be as accurate as possible.

\subsection{The ADCIRC Model}

Storm surge is a physical process that is a result of sea water flow induced by winds and is often further increased in magnitude by tides.  In some cases, a distinction is made between storm tide and storm surge, with storm surge referring only to the excess water elevation above the natural tide. We do not make this distinction. By storm surge, we mean the total water elevation above the datum (e.g. what is sometimes called storm tide). 
The governing model we use for shallow water flow are the two dimensional  shallow water equations (SWE) which consist of the  depth averaged equations of mass conservation as well as $x$ and $y$ momentum conservation \cite{tan1992shallow}:
\begin{equation} \label{eq:SWE}
\begin{array}{ll}
\quad \text{Find }  (\zeta, \uu)   \text{ such that:}  \qquad \qquad    \\ \\
 \frac{\partial  \zeta}{\partial t} + div (H{\uu})  = 0, \text{ in } \Omega, & \\ \\
\frac{\partial (Hu_x)}{\partial t} + div \left( Hu_x^2 + \frac{g}{2}(H^2-h_b^2), Hu_xu_y \right) - g\zeta \frac{\partial h_b}{\partial x}  = F_x, \text{ in } \Omega, & \\ \\
\frac{\partial (Hu_y)}{\partial t} + div \left( Hu_xu_y, Hu_y^2 + \frac{g}{2}(H^2-h_b^2) \right) - g\zeta \frac{\partial h_b}{\partial y}  = F_y, \text{ in } \Omega, 
 \end{array}
\end{equation}
where $\zeta$ is the free surface elevation above the geoid,  $h_b$ the bathymetry, $H =\zeta + h_b$ is the total water column,  $\uu = \{ u_x,u_y\}^{\text{T}}$ is the depth averaged velocity field, and the source terms $F_x,F_y$ represent potential relevant sources which induce flow, including: Coriolis force, tidal potential forces, wind stresses, and wave radiation stresses. $\Omega$ is the computational domain, e.g., the coastal ocean. To solve this set of transient nonlinear partial differential equations for physically relevant shallow water flows in storm surge, a numerical method is required. In this study, we consider the well established numerical shallow water equation solver ADCIRC~\cite{luettich1992adcirc,pringle2021global}. ADCIRC was initially developed by Luettich, Westerink, and Scheffner to model shallow water flows in coastal regions, estuaries, and shelves.

ADCIRC uses a continuous Galerkin finite element method to spatially discretize the weak form corresponding to~\eqref{eq:SWE} and utilizes a finite difference technique to advance the solution in time.  Over the last few decades, ADCIRC has been extensively developed and improved to include physics and other features that are critical to accurate modeling of shallow water flows during hurricane storm surge events including: waves, tides, bottom friction, levees and floodwalls, wetting and drying, and high resolution representation of bathymetry and topography. ADCIRC distinguishes itself from many other coastal circulation models in the use of the finite element method for spatial discretization. This allows users to develop unstructured meshes of variable resolution throughout the computational domain. Hence, it is well suited to model phenomena such as hurricane storm surge which relies on both computational efficiency as well as variable resolution of a large domain, e.g., the Atlantic Ocean and the US East Coast. Furthermore, ADCIRC has been optimized for high-performance computing through MPI parallelization and is the backbone of several operational storm surge forecasting systems, see, e.g., Dresback \emph{et al.}~\cite{dresback2013skill}.  ADCIRC has been extensively validated for past hurricanes, including Ike~\cite{hope2013hindcast}, Gustav~\cite{forbes2010retrospective} (both 2008), Katrina, Rita (both 2005)~\cite{dietrich2012performance}, Harvey (2017)~\cite{goff2019outflow}, as well as others. Due to its extensive validation and capabilities, ADCIRC represents an ideal candidate to generate training data for our surrogate models.

\subsection{Texas}

The first region we will consider is the coast of Texas. Fortunately, there is a preexisting publicly available dataset of ADCIRC output for 446 synthetic storms that make landfall on the Texas coast~\cite{designsafe2021storms}. The synthetic storms were originally developed to assess flooding risk for computation of insurance premiums. The dataset is similar to the synthetic storm databases used in previous surrogate modeling studies~\cite{lee2021rapid,jia2016surrogate,BASS2018159}.
For each synthetic storm we have the best track data, as well as wind and pressure fields with a .05 degree spatial resolution and a 15 minute temporal resolution. Time-series water elevation output is available for the entire spatial domain at a temporal resolution of two hours, in addition to the the maximum attained surge.

\begin{figure}[h!]
    \centering
    \includegraphics[width=1.0\textwidth]{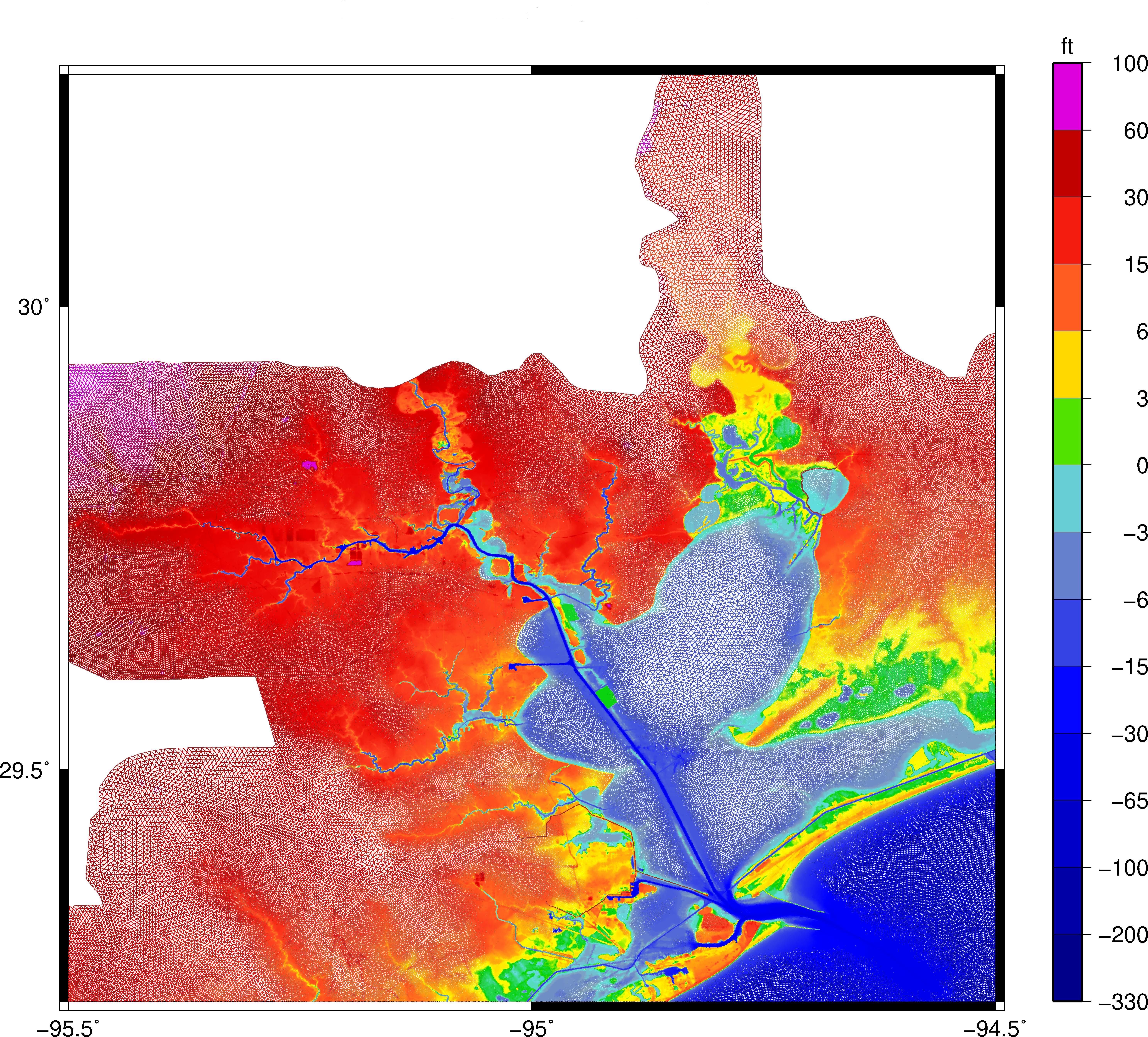}
    \caption{ADCIRC mesh from~\cite{hope2013hindcast} in the Galveston Bay area.}
    \label{fig:mesh}
\end{figure}

%was validated for the Texas coast in~\cite{hope2013hindcast}. It
% the Ike study actualy uses a different mesh . . .
The ADCIRC mesh used to generate the synthetic storms contains $3,352,598$ nodes and $6,675,517$ triangular elements. This mesh covers the entire North Atlantic ocean  and the Gulf of Mexico to ensure necessary resolution of all far-field events physics that are critical to accurately resolve storm surge on the Texas coast. The mesh contains extraordinary high resolution on the Texas coast with element size on the order of 10$m$, see Figure~\ref{fig:mesh}, for a plot of the mesh in Galveston Bay.
To ensure physically relevant simulations of flow near the coast, in estuaries, and in floodplains, this mesh also contains a a spatially variable classification of the sea bottom and land defined though a friction coefficient used in a Manning's $n$ friction formula~\cite{manning1890flow} to ascertain physically relevant friction forcing.

Each storm in the database is a tropical cyclone. Consequently, the database of storms includes best track data as well as meteorological forcing and water elevation output. Best track data consists of the location of the storm's eye, the central pressure, and a number of other storm parameters. In this study, we will only make use of the eye location data, all other storm parameters are ignored.

\subsection{Alaska}

The second region considered in this study is Alaska. An often under-studied region, modeling storm surge adequately along the Alaskan coastline has become increasingly important as it has experienced increasingly frequent and intense Extra-Tropical Cylcones (ETCs) \cite{graham2001evidence, terenzi2014stormsurge}. These events, such as the recent Typhon Merbok (2022), produced surges greater than 3m threatening Alaska's critical and unique natural, social, and economic system and costing millions of dollars in damage \cite{henry1976storm, reimnitz1979effects, kowalik1984storm, johnson1986modeling, kinsman2012coastal, wicks2017identification}.  While many prior studies have constructed surrogate surge models for the Texas coast, to our knowledge this work is the first to do so for Alaska. 

Many challenges exist in modeling storm surge along the Alaskan coast. First, as it is an under-studied region, there is little available data in the sense that there are fewer gauges measuring water elevation values and few, if any, comprehensive data-sets on flooding events in the region \cite{buzard2021coastal}. Furthermore, since ETCs develop sometimes from not just closed low pressure systems that can be easily parameterized, best track data is generally not available for these storms and a good tracking algorithm that captures the full surge event is generally not available \cite{mesquita2009new}. Lastly, there is the additional challenge of modeling the effects of sea ice coverage in the region, as this has great variable and non-trivial impact on the momentum transfer from air to sea as well as propagation of surges \cite{kowalik1984storm, johnson1986modeling, blier1997storm}. This effect is often critical for accurate simulations and is just now being incorporated into models via parameterizations of the sea-ice interaction based on the sea-ice coverage \cite{joyce2019high}. 

To build a training set for storm surge along the Alaska coastline we first had to identify storm surge events. To-do so, we apply a modified version of the identification algorithm used in \cite{wicks2017identification}. The identification algorithm operates by finding intervals where the residual between predicted tides and observed water levels for a given tidal station gauge is above a given threshold, called the trigger threshold. The challenge becomes to cluster these events into discrete non-overlapping events. Our simplified version of \cite{wicks2017identification}'s algorithm can be summarized in Algorithm~\ref{alg:surge}.
\algrenewcommand\algorithmicfunction{}
\begin{algorithm}[h!]
  \caption{Storm Surge Events Algorithm: Algorithm based off of \cite{wicks2017identification}.\\
  $R(t_i): $ The residual between the predicted tides and the observed water levels at times $\{t_i\}_{i=1}^m$\\
  $T : $ Trigger threshold (meters)\\
  $c : $ Continuity factor $0 < c <= 1$\\
  $L, S : $ Lull and Shoulder periods (hours)\\
  }
  \label{alg:surge}
  \begin{algorithmic}[1]
    \Function{\textbf{GET-SURGE-EVENTS}}{$R, T, c, L, S$}
        \State $\{s^j\}  \leftarrow$ Contiguous distinct sets $s^j = \{ t_i : R(t_i) >= T \}$
        \Comment{Identify candidate surge events}
        \State $\{g^j\}  \leftarrow$ Contiguous distinct sets $g^j = \{ t_i : R(t_i) < T \}$
        \Comment{Time gaps between events.}
        \For{$g^j$}
            \If{If $R(t_i) >= cT \forall t_i \in g^j$ OR $\text{len}(g^j) < L$}
                \State $s^{j-1} \leftarrow s^{j-1} + s^j$
                \Comment{Join events around $g^j$ if $cT$ not crossed or gap shorter than $L$}
            \Else
                \State $s^j \leftarrow \{t_i: t_i > \text{min}(s^j) - S \} \cup s^j \cup \{t_i: t_i < \text{max}(s^j) + S\}$
                \Comment{Add $S$ hours before/after event}
                \State $E \leftarrow s^j $
                \Comment{Add to identified events set}
            \EndIf
        \EndFor
        \State \Return $E$
    \EndFunction
  \end{algorithmic}
\end{algorithm}
\begin{figure}[h!]
    \centering
    \includegraphics[width=1.0\textwidth]{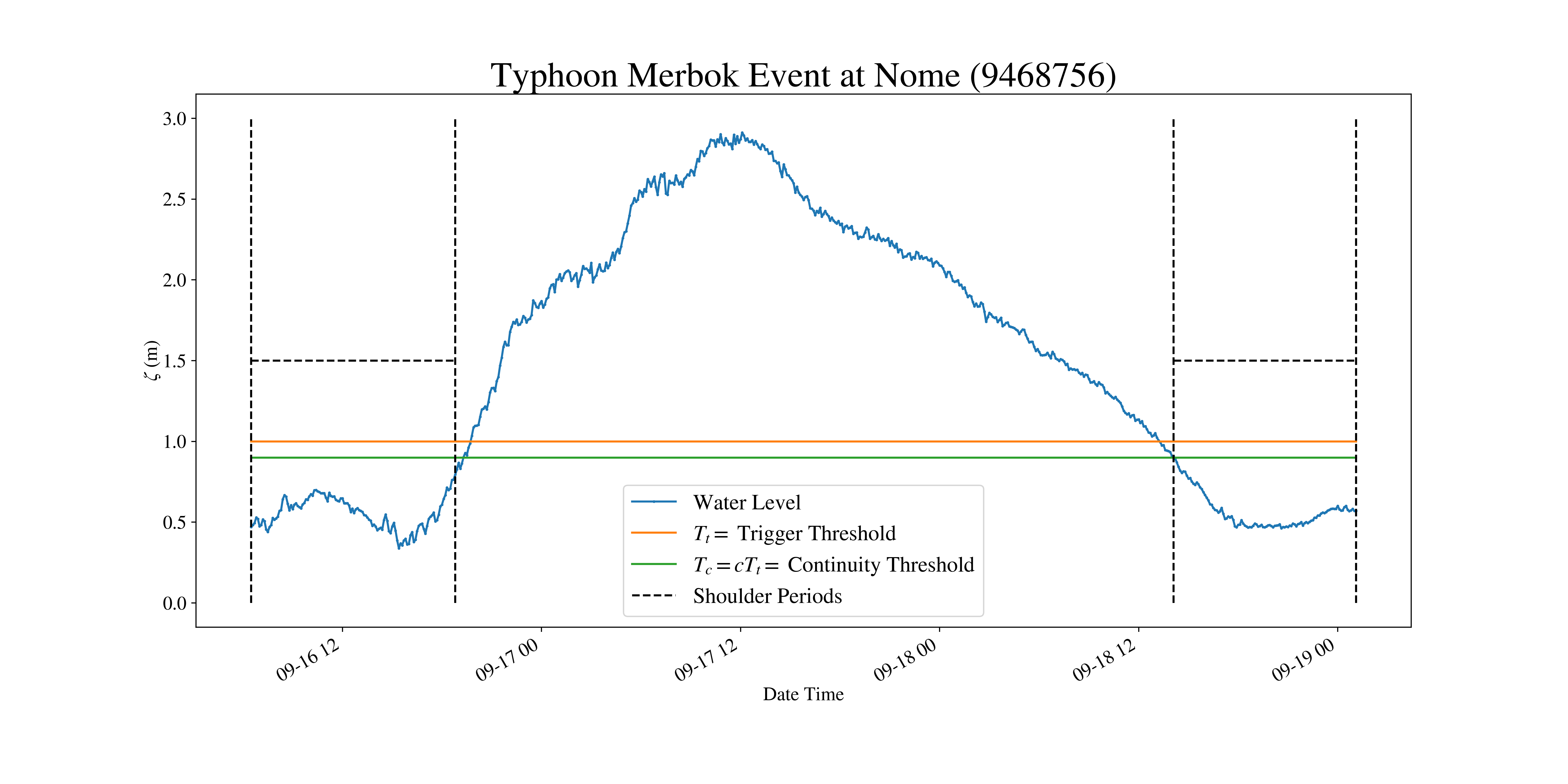}
    \vspace{-1.5cm}
    \caption{Typhoon Merbok event as identified by Algorithm \ref{alg:surge} at Nome, Alaska}
    \label{fig:merbok_nome}
\end{figure}
We used the publicly available NOAA CO-OPs API \cite{NOAATidesCurrents} as our source for tidal gauge data, and focused on the three stations that see the most frequent storm surge events: Nome, Red Dog Dock, and Unalakleet. We ran Algorithm \ref{alg:surge} for each storm, selected only the positive surge events, and then combined the date-ranges across each station, merging overlapping events. See Figure \ref{fig:merbok_nome} as an example of how the algorithm identified the peak surge event associated with Typhoon Merbok at Nome Alaska. 

Having identified date-ranges of relevant storm surge events, we compiled meteorological forcing data from the CFSv1 \cite{saha2006ncep} and CFSv2 \cite{saha2014ncep} data-sets, including wind, air pressure, and ice concentration. A lightweight version of the ADCIRC mesh used in \cite{joyce2019high} consisting of 443,770 nodes and 850,855 triangular elements  was used and ADCIRC was run to compute the true water levels across the whole domain for each storm surge event and compile the final raw training data-set. We note once again, the the base training data-set consist of just the meteorological forcing data (wind, pressure, and sea ice coverage) and ADCIRC output water levels, with no best track data (unlike Texas). The base data-set of storm surge events and ADCIRC simulations has been published to the DesignSafe \cite{rathje2017designsafe} Data repository \cite{del-castillo-negrete2023alaska}.

\section{Model Construction}
\label{sec:model}

Let $\Omega$ denote the spatial computational domain with coordinates $\mathbf{x} \in \Omega$, and let $t \in \mathbb{R}$ denote time. A storm $s$ is defined by vector valued meteorological forcing data $\mathbf{f}_s(\mathbf{x}, t)$ (e.g.,  wind and pressure). The corresponding storm surge is given by the scalar valued function $\eta_s(\mathbf{x}, t)$ and is presumed to functionally depend on the forcing $\mathbf{f}_s$. The time dependent storm surge prediction problem computes the mapping from $\mathbf{f_s}$ to $\eta_s$. While ADCIRC and other high-fidelity numerical models solve the time dependent storm surge prediction problem, here we solve the simplified problem of predicting maximum storm surge, defined as $\eta^{max}_s(\mathbf{x}) = \max_t\eta_s(\mathbf{x}, t)$. In many applications, such as coastal risk assessment, knowledge of $\eta^{max}$ is sufficient. Thus, we will drop the superscript and refer to the maximum storm surge as $\eta(\mathbf{x})$. 

\subsection{Point-based formulation}
\label{sec:surrogate_formulation}

Constructing a surrogate model requires a set of sample inputs and outputs from a high-fidelity storm surge model, e.g $(\mathbf{f}_s(\mathbf{x}, t), \eta_s(\mathbf{x}))_{s=1}^{n}$. Because of computational constraints, $n$ is typically small, a few hundred or thousand. However, replicating the high resolution outputs of storm surge model requires making predictions at a large number of points $N$. Directly predicting a high dimensional output vector with a model such as a neural network would require a large number of model parameters relative to the number of training examples. Such a model could be impossible to train due to data limitations (i.e the number of surge events in the training dataset would be much smaller than the output dimension). Consequently, a different approach is needed.

One solution is to predict the storm surge only at a small number of points, as is often done in the literature \citep{LEE2006483,kim2015time}.  
However, this approach is problematic if high resolution output is desired, as interpolation from the sparse predictions can result in significant error. Another approach is to use dimensional reduction techniques such as principal component analysis~\cite{jia2013kriging, jia2016surrogate, kyprioti2021storm}. However, this explicitly constraints the model to a low dimensional manifold generated by the training data. In the case of principal component analysis, the manifold is linear. This approach could potentially struggle to capture the full range of physical outcomes, as there is no guarantee that the space of potential maximum surge profiles can be effectively projected into a low-dimensional linear space.
 
To circumvent the need for dimensional reduction while avoiding a large number of model parameters, we propose a novel \textit{point-based} formulation of the surrogate problem. Instead of considering one storm $s$ as one training example, we consider each combination of storm and output location $(s, \mathbf{x_i})$ as a training example. While this significantly increases the number of predictions that are needed, it reduces the model output dimension to 1. This reduces the minimum number of model parameters, while simultaneously increasing the number of training examples. Both of these changes make the estimation problem more computationally feasible. Another added benefit of this formulation is that the model can directly make predictions for locations it has not seen in the training data, including points potentially outside of the study domain. Furthermore, the point-based formulation gives us the flexibility to make predictions only in a small area of interest within a much larger study domain.

\subsection{Feature Engineering}
The preceding abstract formulation applies to both study areas. The next steps are to construct a good feature representation and select an appropriate machine learning model. These steps will differ somewhat for each study area, as there are differences in the available data. However, the basic approach is the same.

Since the modeling problem is posed for a single point in the domain, the feature representation needs to encode only the information relevant to predicting storm surge for the given location. Although the surge at any given point can depend on the entirety of the bathymetry, wind, and pressure fields, in practice not all of the data will be relevant. Our goal is to develop a compact feature representation that is still robust enough to enable accurate predictions. In addition, we want to discourage overfitting the data. For instance, we exclude the latitude and longitude from the training data. While storm surge does vary significantly by location, this variation is driven by bathymetry and coastline shape, not the actual values of the latitude and longitude.

\subsubsection{Texas}

The Texas dataset includes best track data and consists of synthetic tropical cyclones that make landfall on the coast. This allows us to identify both the landfall location and time for each cyclone. We use this information to localize the input data in space and time. The wind and pressure data are provided as time series, but the output (peak surge) is not time dependent. To remove the temporal component of the data, we first identify the time of landfall from the best track data, and subset the data to a fixed window from six hours before landfall to six hours after landfall. Secondly, we compute the mean, max, and min for the wind vector, the pressure, and the wind magnitude over the subset data. This process produces a total of twelve variables (three temporal statistics each for wind $x$-direction, wind $y$-direction, wind magnitude and air pressure). Note that the synthetic cyclone events in the Texas dataset do not include tidal forcing.

We expect storm surge at a given point to depend on the forcing data and bathymetry in a spatial neighborhood about the point.
To account for this spatial dependency, we compute statistics for each of the twelve forcing variables and for the bathymetry. These are computed for each point over a sequence of neighborhoods that increase in size. Each neighborhood is a regular lat-lon box. For forcing data, the neighborhood sizes are 0.1 degrees, 0.2 degrees, 0.4 degrees, and the entire spatial domain. Neighborhood sizes of 0.05, 0.1, 0.4, and 1.0 degrees are used for the bathymetry.
We compute the mean, max, and min of the underlying variable for each neighborhood.
%Because the ADCIRC mesh used to generate the training data has variable resolution, the data are not gridded and have an irregular spatial distribution. Computing the spatial statistics is much easier when working with regular data, so we first impose a regular .01x.01 degree grid on the spatial domain and take means over each grid cell. Spatial statistics are then computed over the regular grid.    
In addition to spatial and temporal statistics, we add the distance to the landfall location, the distance to the coastline, and the raw bathymetry. This results in 135 features for the Texas data.

%However, many of these are highly correlated and likely redundant. Correlation can also cause stability issues for training neural networks, so it is desireable to reduce the number of highly correlated features. We set a correlation threshold of .9, and removed redundant features until no pairs remained that were correlated above the threshold. This reduces the number of features to 58 - a reduction of nearly 60\%.
% include a list of the features in the appendix - both the full original list and the reduced ones

\subsubsection{Alaska}

For the Alaska dataset, we follow a similar approach as outlined for Texas in the preceding section. However, no best track data is available, and not every event is a cyclone which makes landfall. Consequently, we use all of the forcing data for temporal statistics, and compute the same spatial statistics as before. The ADCIRC forcing inputs for Alaska include sea ice concentration  which doesn't exist in the Gulf of Mexico but can have a significant effect in the Bering Sea. We process the ice concentration the same way as the other forcing inputs by  first  computing temporal and then spatial statistics.

Unlike the Texas dataset, the Alaska dataset consists of real surge events, most of which are not hurricane strength intensity. Furthermore, Alaska experiences much stronger tidal variations than the Gulf Coast. The combination of weaker winds and stronger tides means that tidal forcing will have a greater effect on storm surge. Consequently, it is critical to include tidal forcing in the training data.

In the absence of winds or pressure variations, the tide at a given point on Earth can be modeled to high accuracy by a linear combination of \textit{tidal constituents}. Each tidal constituent is a sinusoidal function. The phase and amplitude for each constituent vary by location and ultimately depend on the bathymetry and coastline. When the phases and amplitudes are known for a given location, tides can be predicted accurately over decades-long time horizons \cite{lynch1979wave}. Given a set of tidal constituents, ADCIRC can compute the corresponding tidal amplitudes and phases for each node in the finite element mesh. We used ADCIRC to determine amplitudes and phases at each node in the Alaska mesh for eight major tidal constituents: M2, S2, N2, K2, O1, K1, P1, and Q1. The amplitudes of the constituents were added as features. 
%While a more complex tidal analysis was certainly possible (for instance, we could have attempted to directly compute tides), we elected for the simpler approach to reduce model complexity. %% I removed this, if we want to we can mention this as potential future work in the conclusions
A total of 172 features are used for the Alaska dataset. All features in the Texas dataset are present with the exception of the distance to landfall.

\subsection{Model Architecture}

One challenge for modeling storm surge is that for any given storm, a large number of points in the spatial domain may experience zero surge. Predicting which locations will remain dry is essentially a classification problem, not a regression. On the other hand, predicting the level of surge for inundated areas is a regression problem. It makes sense to solve each problem separately instead of trying to tackle them simultaneously. Consequently, we construct two models, one for classification of points as wet or dry, and one for predicting the level of inundation at wet points.

For both the classification and regression stages, we consider a variety of model architectures. Because the underlying shallow water equations are nonlinear, storm surge is a nonlinear process. Consequently both the classification of points as inundated or not and the prediction of maximum inundation levels require a model with the ability to approximate nonlinear functions. ANN are capable of capturing highly nonlinear interactions, and have been applied successfully across a vast array of disciplines, making them a popular choice for surrogate models. In this work we consider a small number of neural network architectures, as well as a gradient boosting method that is representative of non-neural methods. 

\begin{figure}
    \centering
    \includegraphics[width=1.0\textwidth, height=60mm]{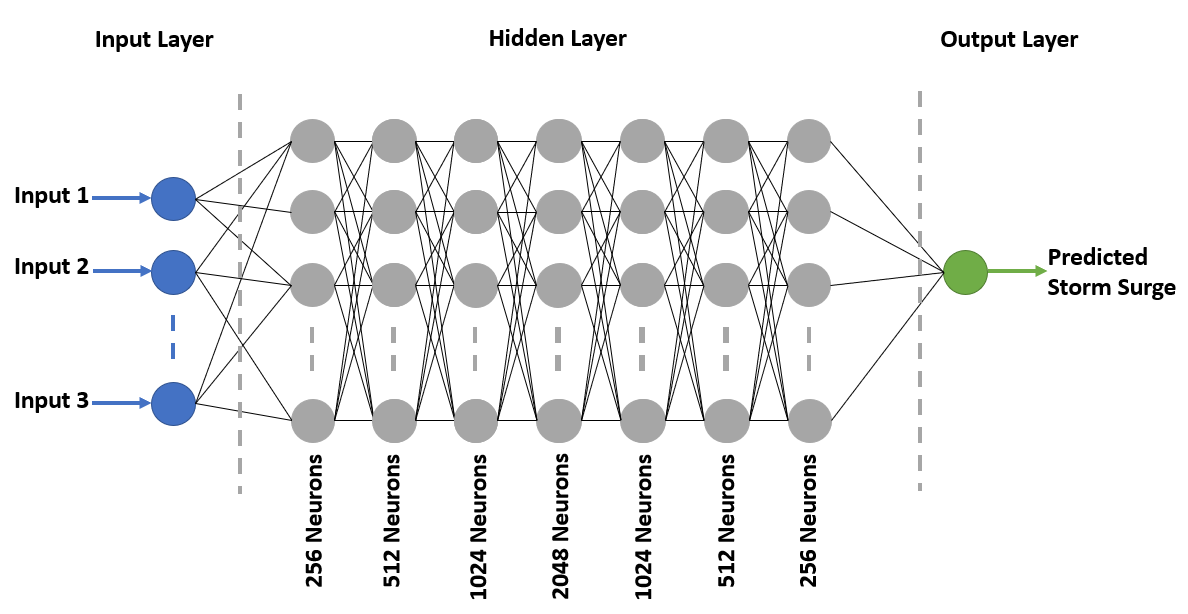}
    \caption{Feed-forward network architecture.}
    \label{fig:net_arch}
\end{figure}

For both the classification and regression stages, we consider a total of three feed-forward network architectures with ReLU \cite{agarap2019deep} activation. The largest of these is illustrated in Figure \ref{fig:net_arch}, and a list of the network architectures is given in Table \ref{tab:nn_arch}. The only difference between the classification and regression networks is the output neuron, sigmoid activation \citep{NARAYAN199769} for classification and ReLU for regression. We remark that the choice of ReLU activation for the regression networks is appropriate because the maximum inundation level for inundated points will be non negative.

\begin{table}[h!]
    \centering
\begin{tabular}{llrr}
\toprule
Neural Network & Number of Hidden Layers &    Size of Hidden Layers \\
\midrule
   nn1 &    3 & (256, 512, 256)  \\
   nn2 &    5 & (256, 512, 1024, 512, 256) \\
   nn3 &    7 & (256, 512, 1024, 2048, 1024, 512, 256) \\
\bottomrule
\end{tabular}
\caption{Architecture of Neural Networks}
\label{tab:nn_arch}
\end{table}

For gradient boosting we use the popular XGBoost library \cite{chen2016xgboost}. XGBoost can be used for both classification and regression problems. The addition of XGBoost brings the number of candidate model architectures up to four for both the regression and classification stages. This results in a total of sixteen possibilities for the full two-stage model.

\subsection{Model Training}

For each dataset, we used 90\% of the surge events for training, and the remainder for testing. All neural network models were trained using a single NVIDIA Ampere A100 GPU on the Lonestar6 system at the Texas Advanced Computing Center. The XGBoost models were trained in parallel on 32 CPUs on Lonestar6. All classification models used a binary cross entropy loss function. The regression models minimized the mean squared error.

\subsubsection{Texas}

The number of epochs was set to 50 for the classification networks and 100 for the regression networks. This was done to balance training times between the two stages, as more data points are used in classification than regression (regression only uses inundated locations). However, convergence occurs within 10 or so epochs for both stages. The learning rate was set to $10^{-4}$ and adaptively decreased with a minimum of $10^{-6}$. The Adam optimizer \cite{reddi2019convergence} was used for both stages. 
Training for all neural networks required less than a half hour for each network. The XGBoost model used 250 boosting rounds and had a comparable training time to the neural network models. 

While the  ADCIRC mesh used for the Texas study consists of over 3 million nodes, only a subset were used to train the model for efficiency. We performed several reductions: first, predictions are only made for coastal points that are within a radius of 150 km of the landfall location. This was done to focus the model's predictions on regions that experience significant surge and avoid a bias towards underprediction of the most extreme surges. Second, after applying these restrictions, we further reduce the number of points by a factor of 10. This allows the training dataset to fit into RAM and makes training times reasonable. Preliminary tests showed that using larger numbers of points did not improve the model accuracy, while significantly increasing training time.

\subsubsection{Alaska}

The mesh used for the Alaska data was much smaller, with just over 400K nodes.  In addition, there were only 109 events in the training dataset, as opposed to 447 in the Texas dataset. We still restrict predictions to coastal points, but do not perform further reductions. Training times were significantly shorter than for the Texas models with around five minutes for both the XGBoost model and neural networks. The same hyperparameters were used. The main difference is that the neural network models converge within an epoch or two, the train loss keeps improving, but the validation loss stops improving.

Because the Alaska data consists of real surge events, we partitioned it temporally instead of randomly. The surge events are first sorted by event start date. Data from the first 90\% of the events is used for training. The remaining 10\% of the data correspond to events that take place after those in the training set. Specifically, the test data include the recent Typhoon Merbok, an important test case for validation of the model.

\section{Results}
\label{sec:results}

\subsection{Texas}

We began by determining the optimal choices of model for the classification and regression stages via grid search and the results are summarized in Table \ref{tab:texas_grid}. We trained a total of sixteen two-stage models. Overall the variability in performance was low. The best performing model used the largest neural network for both stages, and had a root mean squared error (RMSE) of 0.31 meters. The worst performing model used XGBoost for both stages, and had an RMSE of 0.36 meters. Clearly, the model performance improves as the number of parameters increases. However, the improvements in performance become increasingly marginal as the model complexity grows. Using a larger network than nn3 did not significantly improve performance.

We remark that the error metrics are computed across all points, including those whose inundation is incorrectly classified by the model. Consequently, the metrics reflect the performance of both stages of the model.

\begin{table}[h!]
    \centering
\begin{tabular}{llrr}
\toprule
classifier & regressor &    $R^2$ &  RMSE \\
\midrule
   xgboost &   xgboost & 0.829 & 0.355 \\
   xgboost &       nn1 & 0.850 & 0.333 \\
   xgboost &       nn2 & 0.852 & 0.330 \\
   xgboost &       nn3 & 0.853 & 0.329 \\
       nn1 &   xgboost & 0.838 & 0.345 \\
       nn1 &       nn1 & 0.850 & 0.333 \\
       nn1 &       nn2 & 0.857 & 0.325 \\
       nn1 &       nn3 & 0.857 & 0.325 \\
       nn2 &   xgboost & 0.845 & 0.338 \\
       nn2 &       nn1 & 0.869 & 0.311 \\
       nn2 &       nn2 & 0.865 & 0.316 \\
       nn2 &       nn3 & 0.870 & 0.310 \\
       nn3 &   xgboost & 0.854 & 0.328 \\
       nn3 &       nn1 & 0.866 & 0.314 \\
       nn3 &       nn2 & 0.869 & 0.311 \\
       nn3 &       nn3 & 0.873 & 0.307 \\
\bottomrule
\end{tabular}
\caption{Model performance on the Texas test data.}
\label{tab:texas_grid}
\end{table}

\begin{figure}[h!]
    \centering
    \includegraphics[width=100mm, height=55mm]{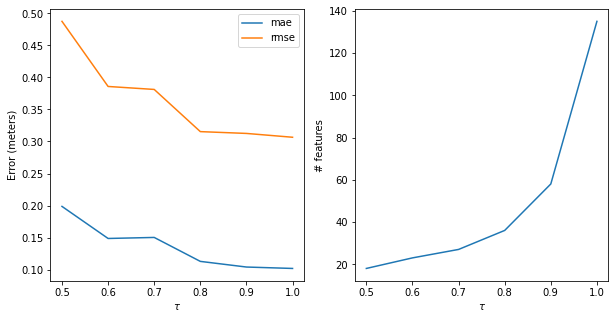}
    \caption{Mean Absolute Error (mae) and Root Mean Squared Error (rmse) with respect to correlation filter threshold.}
    \label{fig:tx_corr}
\end{figure}

In the next experiment, we evaluate the effect of feature reduction on model performance. Many of the features are highly correlated which can result in potential instabilities in neural network training as well as overfitting. In addition, removing correlated features can significantly speed up training time while not impacting performance. To remove correlated features, we set a threshold $\tau$, and iteratively remove features until no pairs of features remain for which the absolute value of the correlation coefficient is greater than $\tau$. This is illustrated in Figure~\ref{fig:tx_corr}, where effect of $\tau$ on the number of features and model performance is shown.    
We find that increasing the number of features improves the performance of the model but the gains are almost negligible when $\tau \ge 0.9$. Consequently the choice of $\tau=0.9$ represents a good trade-off between feature reduction and model performance. The remainder of the presented results for the Texas model are obtained for the model trained on a reduced set of features with $\tau=0.9$. For lists of the full and reduced feature sets, see Table \ref{tab:feature_list}.
% should we add an appendix here?

\subsubsection{Performance on test data}

\begin{figure}[h!]
    \centering
    \includegraphics[width=100mm, height=75mm]{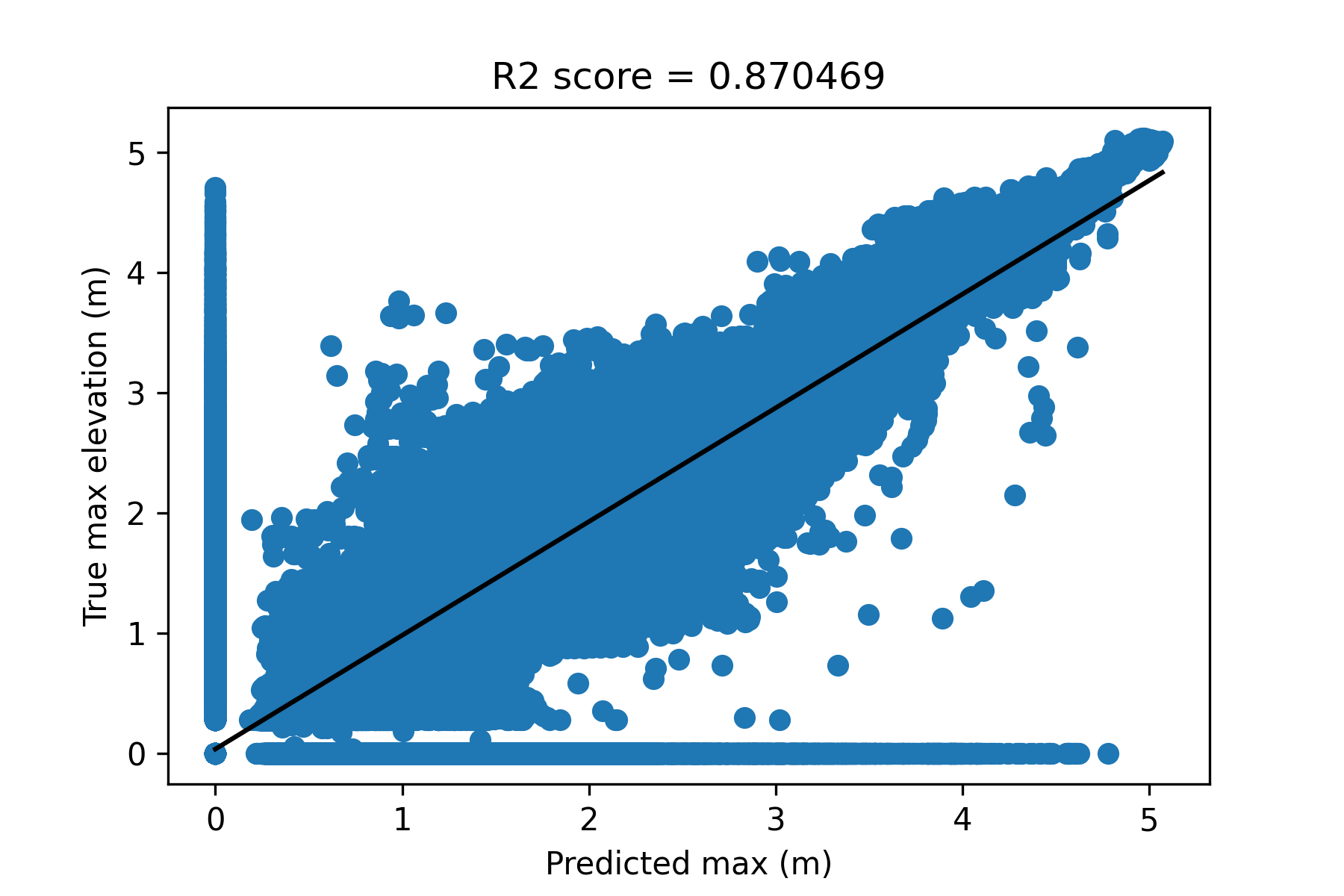}
    \caption{Texas model fit.}
    \label{fig:tx_fit}
\end{figure}

\begin{figure}[h!]
    \centering
    \includegraphics[width=100mm, height=75mm]{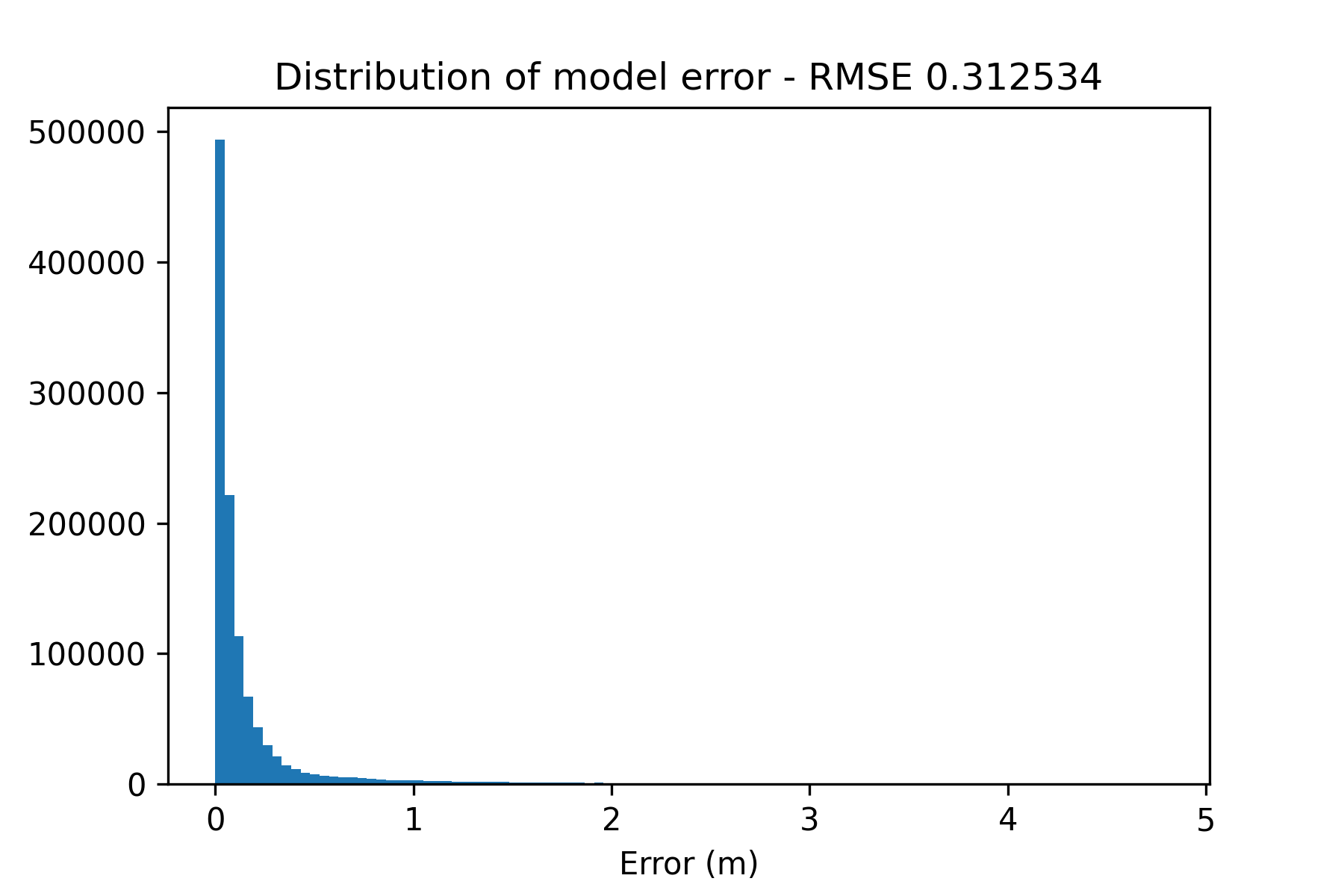}
    \caption{Texas model error distribution.}
    \label{fig:tx_dist}
\end{figure}

\begin{figure}
    \centering
    \includegraphics[width=100mm, height=75mm]{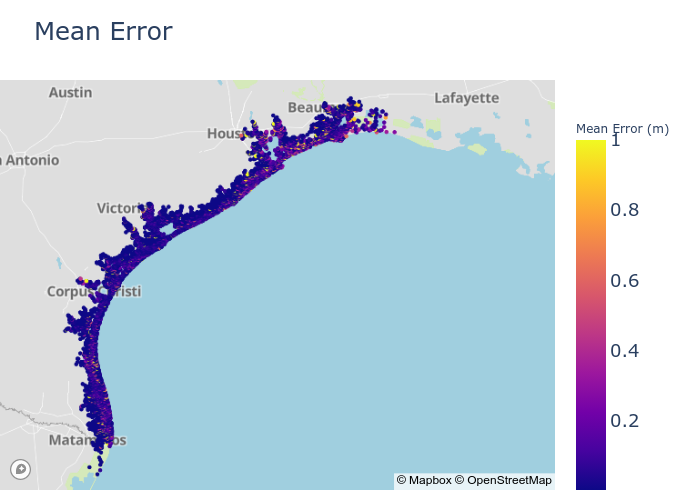}
    \caption{Spatial variation in Texas model errors. Map data courtesy of OpenStreetMap~\cite{OpenStreetMap}.}
    \label{fig:texas_mean_err}
\end{figure}

We find the Texas model achieves an excellent fit on the test synthetic storms  with an R2 score of 0.87 and a RMSE of 0.31 meters (see Figures~\ref{fig:tx_fit} and~\ref{fig:tx_dist}).
To get an idea of the spatial variation in model performance, we compute the average absolute error at each prediction point across all storms, as shown in Figure~\ref{fig:texas_mean_err}. 
In this figure,  we observe that the model performs better on the western portion of the Texas coast, and struggles more around Louisiana. However, the performance is overall quite consistent across the spatial domain.

\subsubsection{Validation on historical storms}

While good performance on synthetic data is encouraging, the ideal test of the model is its ability to replicate ADCIRC's performance for real storms. We performed two tests,  for Hurricanes Ike (2008) and Harvey (2017). For each storm, we ran an ADCIRC simulation on the same mesh used to generate the synthetic data. Each simulation used 740 CPUs on Lonestar6 and completed in around two hours. Following the construction of the training data, we determined the landfall location for each hurricane and predicted maximum surge within a 150km neighborhood of the landfall. No further downsampling was applied. Consequently, the model makes predictions for many locations not present in the training dataset.

\begin{figure}[h!]
    \centering
    \includegraphics[width=100mm, height=75mm]{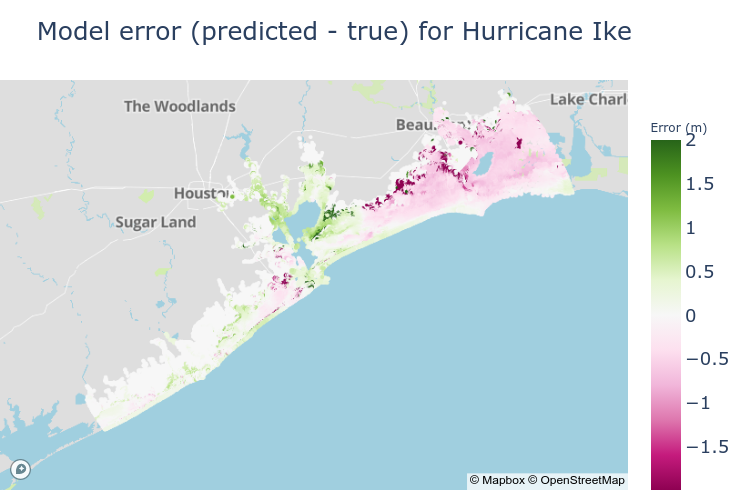}
    \caption{Model error for Ike. Map data courtesy of OpenStreetMap~\cite{OpenStreetMap}.}
    \label{fig:ike_err}
\end{figure}

\begin{figure}[h!]
    \centering
    \includegraphics[width=100mm, height=75mm]{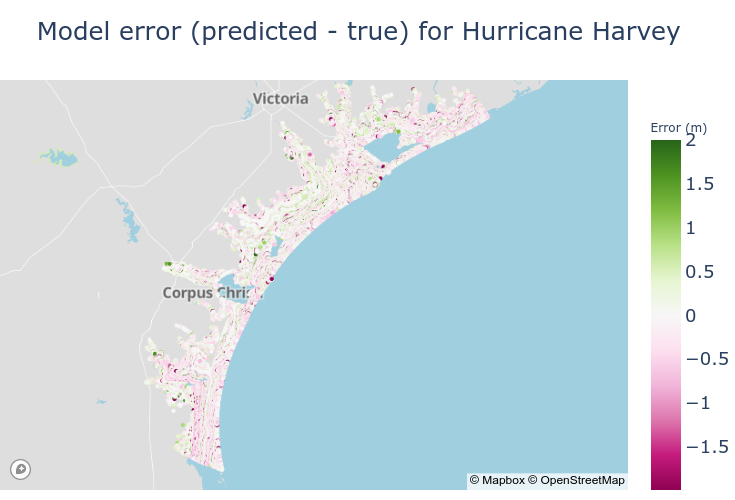}
    \caption{Model error for Harvey. Map data courtesy of OpenStreetMap~\cite{OpenStreetMap}.}
    \label{fig:harvey_err}
\end{figure}

The accuracy is a little lower than the test dataset, but still high. For Ike, the model replicates ADCIRC's predictions with an R2 score of 0.85 and an RMSE of 0.57 meters. For Harvey, the model attains an R2 score of 0.68 and an RMSE of 0.47 meters. Plots of the errors for each storm are shown in Figures~\ref{fig:ike_err} and~\ref{fig:harvey_err}.

\begin{figure}
    \centering
    \includegraphics[width=100mm, height=75mm]{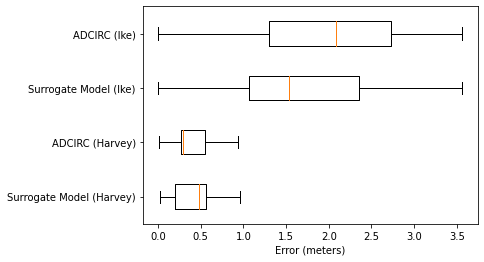}
    \caption{Distribution of errors across observation points.}
    \label{fig:historical_hindcast_boxplot}
\end{figure}

The final test of the model is comparison against actual observations. For Ike we validate the model against 66 maximum elevation observations. These are obtained from a mixture of short-term USGS monitoring devices \cite{east2008monitoring} and NOAA gauges. For Harvey we use 22 maximum elevation observations at NOAA gauges. We compare both the ADCIRC predictions and model predictions to the observed maximum surges. The results are displayed in Figure~\ref{fig:historical_hindcast_boxplot}. 
From the results in this figure, the surrogate performs very similarly to ADCIRC when compared to real observations. Evaluation of our surrogate model takes seconds on a single CPU  whereas running ADCIRC requires hours on hundreds of CPUs. Hence, this surrogate represents a speedup of several orders of magnitude. 
We remark that the ADCIRC performance on Ike is not indicative of the state-of-the-art. Better model performance was attained in \cite{hope2013hindcast} by coupling ADCIRC to a short wave model and running an extensive tidal spinup beforehand. This reflects the fact that the model quality depends on the training data. Nevertheless, we have demonstrated our model is an effective surrogate for the ADCIRC model with no short wave coupling. 

\subsection{Alaska}

%Unlike the Texas dataset, there is little variation in inundation for the Alaska dataset. Consequently, we find no difference in performance for single-stage and two-stage models. The grid search is performed over only the four base model types (XGBoost, nn1, nn2, nn3). We find that the best-performing model is XGBoost - with an rmse of 0.25 meters. The smallest neural network (nn1) attains an rmse of 0.31 meters. Increasing the size of the network results in degraded performance. 

% \begin{table}
% \centering
% \caption{Model performance on the Alaska test data.}
% \label{tab:alaska_grid}
% \begin{tabular}{lrr}
% \toprule
%   model &    R2 &  rmse \\
% \midrule
% xgboost & 0.828 & 0.245 \\
%     nn1 & 0.728 & 0.308 \\
%     nn2 & 0.725 & 0.311 \\
%     nn3 & 0.697 & 0.326 \\
% \bottomrule
% \end{tabular}
% \end{table}

\begin{figure}[h!]
    \centering
    \includegraphics[width=100mm, height=75mm]{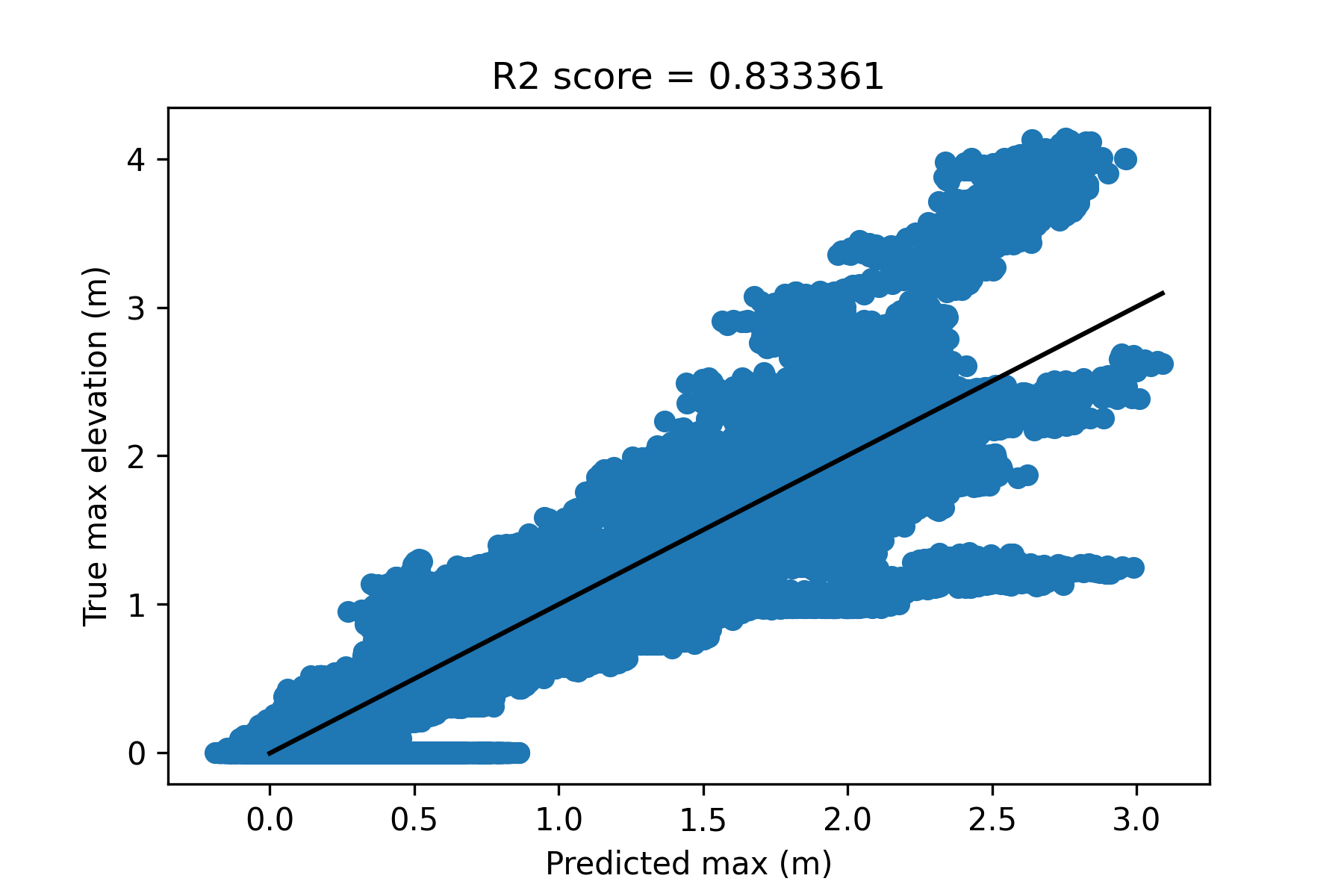}
    \caption{Alaska model fit.}
    \label{fig:ak_fit}
\end{figure}

\begin{figure}[h!]
    \centering
    \includegraphics[width=100mm, height=75mm]{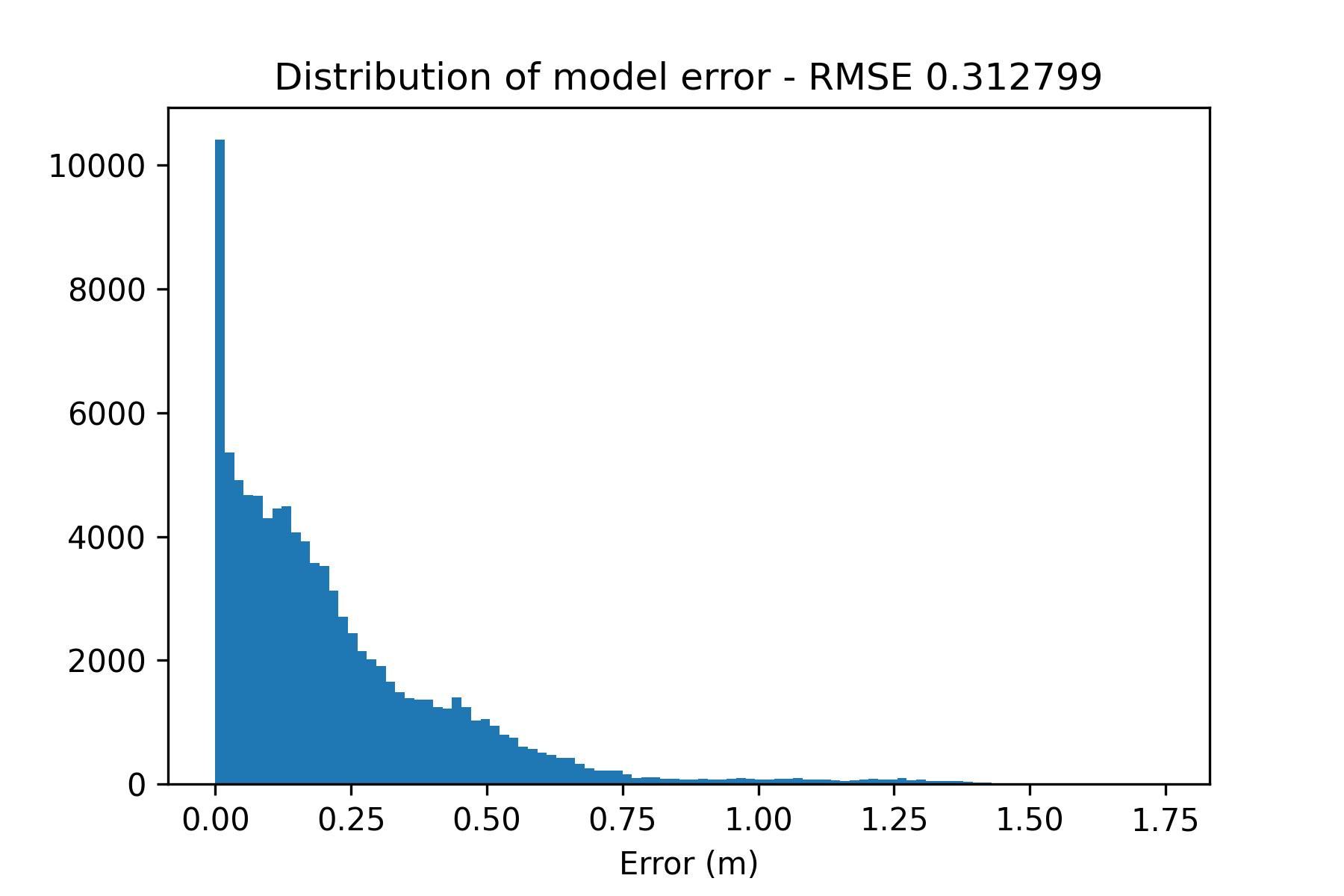}
    \caption{Alaska model error distribution.}
    \label{fig:ak_dist}
\end{figure}

\begin{figure}[h!]
    \centering
    \includegraphics[width=100mm, height=75mm]{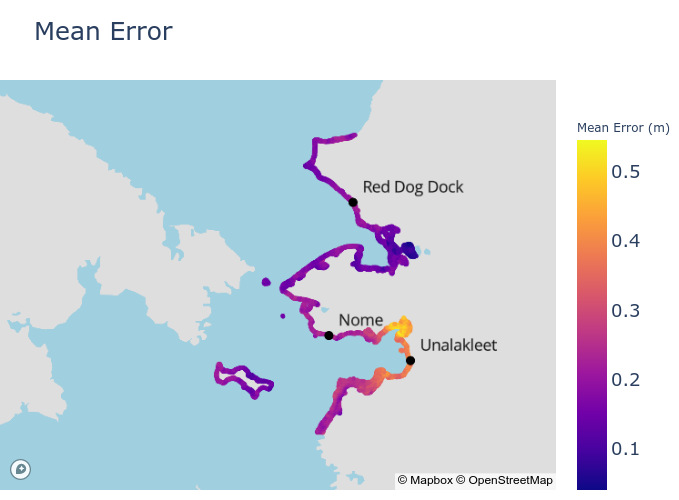}
    \caption{Spatial variation in Alaska model errors. Map data courtesy of OpenStreetMap~\cite{OpenStreetMap}.}
    \label{fig:alaska_mean_err}
\end{figure}

Due to the smaller size of the Alaska dataset, the neural network based models perform significantly worse than gradient boosting. Consequently, we use gradient boosting for both the classification and regression stages. Given the smaller size of the Alaska dataset, there is less performance benefit from dropping correlated features. Furthermore, unlike neural networks, XGBoost does not suffer from instabilities in the presence of correlated features. Consequently, we forgo a feature reduction study, and use all features in the final model. The performance of the XGBoost model on the test dataset is visualized in Figures~\ref{fig:ak_fit} and~\ref{fig:ak_dist}. Here we observe that the performance is comparable to the Texas model. This is encouraging given the much smaller size of the Alaska dataset. We note that the model exhibits a bias towards underprediction of the most extreme surges. This is due to the inclusion of Typhoon Merbok in the test dataset, an event that generated unprecedented levels of storm surge not seen in the historical events data used to train the model. 
The spatial distribution of errors is shown in Figure~\ref{fig:alaska_mean_err} where we observe that the highest errors are clustered around Nome, AK, while the northern portion of the domain has lower error.

\begin{figure}
    \centering
    \includegraphics[width=100mm, height=75mm]{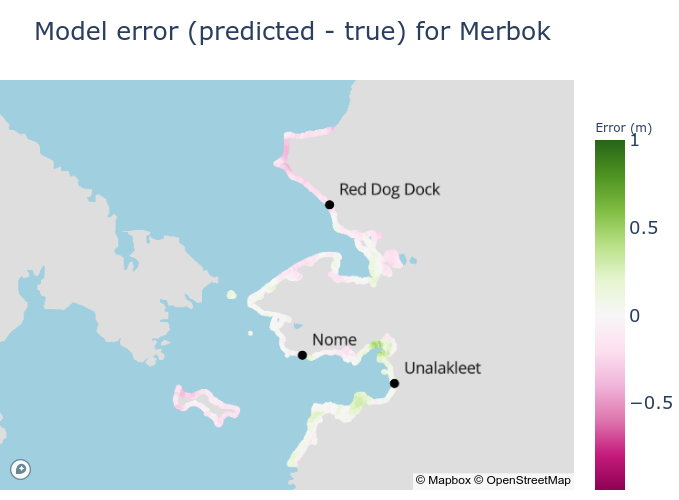}
    \caption{Model error for Merbok. Map data courtesy of OpenStreetMap~\cite{OpenStreetMap}.}
    \label{fig:merbok_err}
\end{figure}

The main reason for the temporal partition of the Alaska data was to ensure that Typhoon Merbok would lie in the test dataset. We find that our model reasonably matches ADCIRC's predictions for Merbok (see Figure \ref{fig:merbok_err}), with an RMSE of 0.59 meters, and an R2 score of 0.93. The RMSE is worse than for the entire test dataset, whereas the R2 score is better. The model underpredicts the highest surge levels for Merbok, but still accurately identifies the locations of maximum surge. The underprediction is unsurprising given that the training data doesn't have a historical event matching Merbok's intensity. 

\begin{figure}[h!]
    \centering
    \includegraphics[width=100mm, height=75mm]{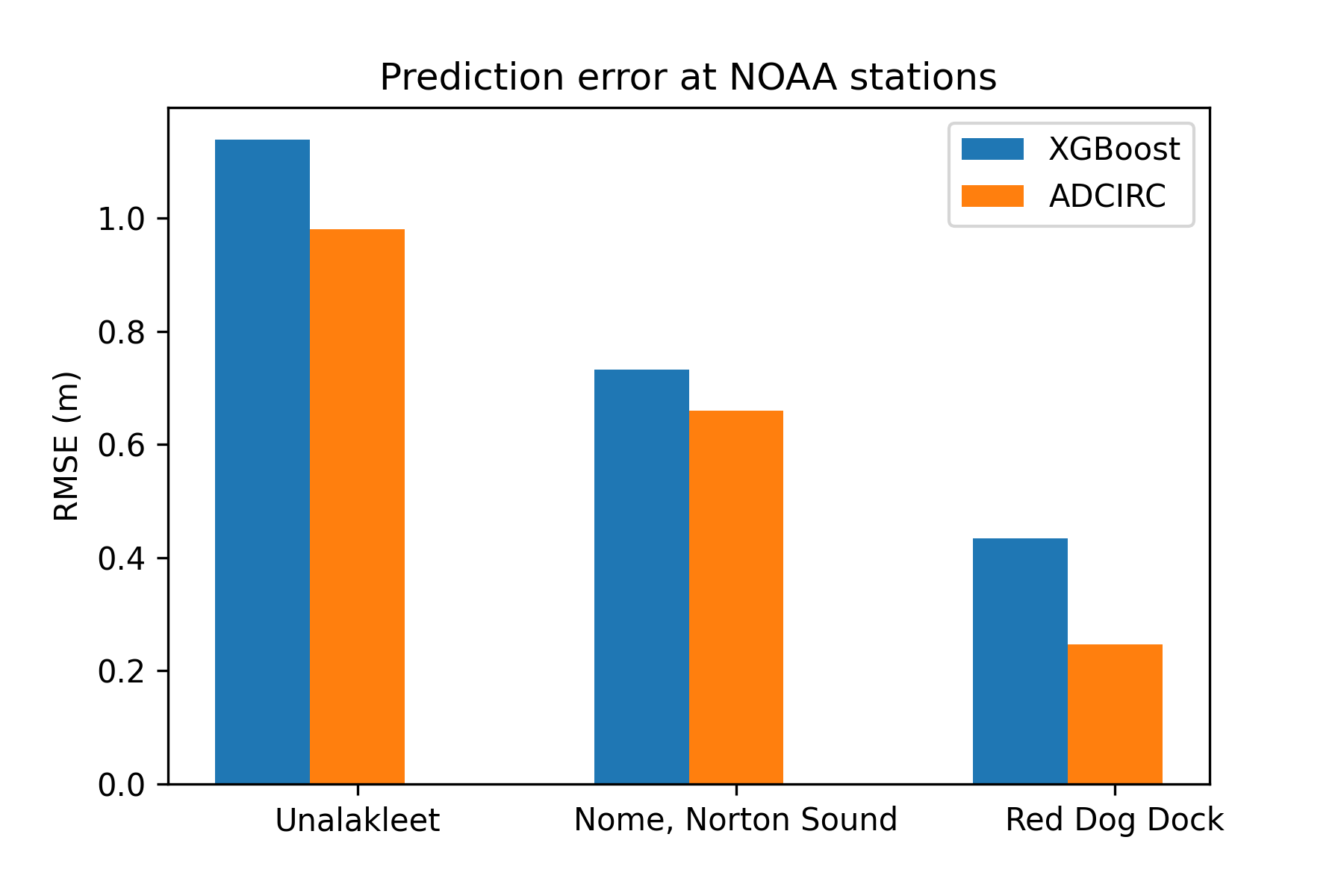}
    \caption{Validation of surrogate model and ADCIRC predictions against NOAA observations. The RMSE is taken over events in the test data.}
    \label{fig:ak_station_error}
\end{figure}

As previously noted, observational data is more sparse for Alaska than for Texas. The only NOAA stations in the study domain are Unalakleet, Nome, and Red Dog Dock. These are the same stations used to select the surge events for the Alaska dataset. For each event in the test dataset, we compare the predictions of the surrogate XGBoost model and ADCIRC to observed NOAA data. At each station, we compute the RMSE over all events in the test data. These results are summarized in Figure~\ref{fig:ak_station_error} where we observe that the ADCIRC model performs slightly better than the surrogate model  with the difference in performance ranging from $0.1$ to $0.3$ meters across stations. Again, the surrogate model has only slightly worse performance than ADCIRC  and is orders of magnitude faster. 

\section{Conclusions}
\label{sec:conclusion}

We have presented a novel framework for developing surrogate models of storm surge. We observe good accuracy for two distinct regions with vastly different training data. The resulting surrogate models use a local formulation that enables them to make predictions at locations not seen in the training data. Our framework is flexible enough to handle both landfalling tropical cyclones and general surge events. In the first case, we were able to accurately predict hurricane surge from synthetic training data. In the second, our model is able to use past events to make accurate predictions for future surge events. The the same pipeline and code were used for these disparate applications  with only minimal adjustments. Comparisons to observational data show that our surrogate models attain similar accuracy as the base ADCIRC model while being orders of magnitude faster.

An interesting direction of future research is the extent to which models trained on one geographic area can be transferred to another. As our surrogate models are not explicitly constrained to a fixed set of points, and don't take in latitudes or longitudes, this is at least a possibility. It may be possible to develop a single, global model of storm surge that subsumes all regional models into one. Generating training data with coverage for the entire globe is infeasible, consequently some sort of transferability is required. While our feature representation allows for good local models, it is still unclear whether it will suffice to enable transferability between completely disjoint regions. At the very least, our local formulation is a step towards full model transferability which is impossible with the traditional approach.

\section*{Acknowledgements}

The authors would like to gratefully acknowledge the use of the "ADCIRC" and "DMS21031" allocations at the Texas Advanced Computing Center at the University of Texas at Austin, as well the use of the DesignSafe platform for data storage.

Funding: This work was supported by the United States Department of Energy under grant DE-SC0022211 - MuSiKAL: Multiphysics Simulations and Knowledge discovery through AI/ML technologies. In addition, Benjamin Pachev was partially supported by a CSEM Fellowship from the Oden Institute for Computational Science and Engineering. The DesignSafe project is financially supported by the National Science Foundation under grant CMMI-1520817.

\appendix

\section{Supplementary Tables}\label{secA1}

Table \ref{tab:feature_list} contains a full list of the features used in each model. Here we give a brief explanation of the feature names. 
\begin{itemize}
    \item Features beginning with `amplitude' indicate a tidal constituent amplitude. The name of the constituent is included as a suffix.
    \item A prefix of `max', `min', or `mean' indicates a temporal statistic - so `max\_pressure' is the maximum pointwise pressure in time.
    \item A suffix of `max', `min', or `mean' followed by a number indicates a spatial statistic. The number denotes the window (in degrees) over which the statistic is computed. For example, `bathy\_max\_0.1' would be the maximum bathymetry over a 0.1 degree window containing the point at its center.
    \item Temporal statistics are computed first - spatial statistics secondly. For instance, `min\_magnitude\_mean\_0.05' is derived by first computing the minimum wind magnitude in time, and then taking a spatial mean.
    \item We note that `iceaf' revers to the fraction of ice coverage, `magnitude' the wind magnitude, `windx' and `windy' the x and y components of the wind, and `pressure' the sea level pressure.
    \item The `depth' feature refers to pointwise bathymetry. The features `landfall\_dist' and `coastal\_dist' refer to distances to the landfall location (for events with best-track data) and the coastline respectively.
    
\end{itemize}

\begin{longtable}{| p{.40\textwidth} | p{.20\textwidth} |p{.20\textwidth} |p{.20\textwidth} |} 
%\centering
\caption{Feature list}
\label{tab:feature_list}
%\begin{tabular}{llll}
%\toprule
\\
\hline
                Feature & Full Texas Model & Reduced Texas Model & Alaska Model \\
%\midrule
\hline
           amplitude\_K1 &                  &                     &            x \\
           amplitude\_K2 &                  &                     &            x \\
           amplitude\_M2 &                  &                     &            x \\
           amplitude\_N2 &                  &                     &            x \\
           amplitude\_O1 &                  &                     &            x \\
           amplitude\_P1 &                  &                     &            x \\
           amplitude\_Q1 &                  &                     &            x \\
           amplitude\_S2 &                  &                     &            x \\
         bathy\_max\_0.05 &                x &                   x &            x \\
          bathy\_max\_0.1 &                x &                   x &            x \\
          bathy\_max\_0.4 &                x &                   x &            x \\
          bathy\_max\_1.0 &                x &                   x &            x \\
        bathy\_mean\_0.05 &                x &                   x &            x \\
         bathy\_mean\_0.1 &                x &                     &            x \\
         bathy\_mean\_0.4 &                x &                   x &            x \\
         bathy\_mean\_1.0 &                x &                     &            x \\
         bathy\_min\_0.05 &                x &                   x &            x \\
          bathy\_min\_0.1 &                x &                   x &            x \\
          bathy\_min\_0.4 &                x &                   x &            x \\
          bathy\_min\_1.0 &                x &                   x &            x \\
           coastal\_dist &                x &                   x &            x \\
                  depth &                x &                   x &            x \\
          landfall\_dist &                x &                   x &              \\
              max\_iceaf &                  &                     &            x \\
      max\_iceaf\_max\_0.1 &                  &                     &            x \\
      max\_iceaf\_max\_0.2 &                  &                     &            x \\
      max\_iceaf\_max\_0.4 &                  &                     &            x \\
     max\_iceaf\_mean\_0.1 &                  &                     &            x \\
     max\_iceaf\_mean\_0.2 &                  &                     &            x \\
     max\_iceaf\_mean\_0.4 &                  &                     &            x \\
      max\_iceaf\_min\_0.1 &                  &                     &            x \\
      max\_iceaf\_min\_0.2 &                  &                     &            x \\
      max\_iceaf\_min\_0.4 &                  &                     &            x \\
          max\_magnitude &                x &                   x &            x \\
  max\_magnitude\_max\_0.1 &                x &                   x &            x \\
  max\_magnitude\_max\_0.2 &                x &                     &            x \\
  max\_magnitude\_max\_0.4 &                x &                     &            x \\
 max\_magnitude\_mean\_0.1 &                x &                     &            x \\
 max\_magnitude\_mean\_0.2 &                x &                     &            x \\
 max\_magnitude\_mean\_0.4 &                x &                     &            x \\
  max\_magnitude\_min\_0.1 &                x &                   x &            x \\
  max\_magnitude\_min\_0.2 &                x &                   x &            x \\
  max\_magnitude\_min\_0.4 &                x &                   x &            x \\
           max\_pressure &                x &                   x &            x \\
   max\_pressure\_max\_0.1 &                x &                     &            x \\
   max\_pressure\_max\_0.2 &                x &                     &            x \\
   max\_pressure\_max\_0.4 &                x &                     &            x \\
  max\_pressure\_mean\_0.1 &                x &                   x &            x \\
  max\_pressure\_mean\_0.2 &                x &                   x &            x \\
  max\_pressure\_mean\_0.4 &                x &                   x &            x \\
   max\_pressure\_min\_0.1 &                x &                   x &            x \\
   max\_pressure\_min\_0.2 &                x &                   x &            x \\
   max\_pressure\_min\_0.4 &                x &                   x &            x \\
              max\_windx &                x &                   x &            x \\
      max\_windx\_max\_0.1 &                x &                     &            x \\
      max\_windx\_max\_0.2 &                x &                     &            x \\
      max\_windx\_max\_0.4 &                x &                   x &            x \\
     max\_windx\_mean\_0.1 &                x &                     &            x \\
     max\_windx\_mean\_0.2 &                x &                     &            x \\
     max\_windx\_mean\_0.4 &                x &                     &            x \\
      max\_windx\_min\_0.1 &                x &                   x &            x \\
      max\_windx\_min\_0.2 &                x &                     &            x \\
      max\_windx\_min\_0.4 &                x &                   x &            x \\
              max\_windy &                x &                   x &            x \\
      max\_windy\_max\_0.1 &                x &                     &            x \\
      max\_windy\_max\_0.2 &                x &                   x &            x \\
      max\_windy\_max\_0.4 &                x &                     &            x \\
     max\_windy\_mean\_0.1 &                x &                     &            x \\
     max\_windy\_mean\_0.2 &                x &                     &            x \\
     max\_windy\_mean\_0.4 &                x &                     &            x \\
      max\_windy\_min\_0.1 &                x &                   x &            x \\
      max\_windy\_min\_0.2 &                x &                   x &            x \\
      max\_windy\_min\_0.4 &                x &                   x &            x \\
             mean\_iceaf &                  &                     &            x \\
     mean\_iceaf\_max\_0.1 &                  &                     &            x \\
     mean\_iceaf\_max\_0.2 &                  &                     &            x \\
     mean\_iceaf\_max\_0.4 &                  &                     &            x \\
    mean\_iceaf\_mean\_0.1 &                  &                     &            x \\
    mean\_iceaf\_mean\_0.2 &                  &                     &            x \\
    mean\_iceaf\_mean\_0.4 &                  &                     &            x \\
     mean\_iceaf\_min\_0.1 &                  &                     &            x \\
     mean\_iceaf\_min\_0.2 &                  &                     &            x \\
     mean\_iceaf\_min\_0.4 &                  &                     &            x \\
         mean\_magnitude &                x &                     &            x \\
 mean\_magnitude\_max\_0.1 &                x &                     &            x \\
 mean\_magnitude\_max\_0.2 &                x &                     &            x \\
 mean\_magnitude\_max\_0.4 &                x &                     &            x \\
mean\_magnitude\_mean\_0.1 &                x &                   x &            x \\
mean\_magnitude\_mean\_0.2 &                x &                     &            x \\
mean\_magnitude\_mean\_0.4 &                x &                     &            x \\
 mean\_magnitude\_min\_0.1 &                x &                     &            x \\
 mean\_magnitude\_min\_0.2 &                x &                     &            x \\
 mean\_magnitude\_min\_0.4 &                x &                     &            x \\
          mean\_pressure &                x &                   x &            x \\
  mean\_pressure\_max\_0.1 &                x &                     &            x \\
  mean\_pressure\_max\_0.2 &                x &                     &            x \\
  mean\_pressure\_max\_0.4 &                x &                     &            x \\
 mean\_pressure\_mean\_0.1 &                x &                     &            x \\
 mean\_pressure\_mean\_0.2 &                x &                     &            x \\
 mean\_pressure\_mean\_0.4 &                x &                     &            x \\
  mean\_pressure\_min\_0.1 &                x &                     &            x \\
  mean\_pressure\_min\_0.2 &                x &                     &            x \\
  mean\_pressure\_min\_0.4 &                x &                     &            x \\
             mean\_windx &                x &                   x &            x \\
     mean\_windx\_max\_0.1 &                x &                   x &            x \\
     mean\_windx\_max\_0.2 &                x &                     &            x \\
     mean\_windx\_max\_0.4 &                x &                   x &            x \\
    mean\_windx\_mean\_0.1 &                x &                     &            x \\
    mean\_windx\_mean\_0.2 &                x &                     &            x \\
    mean\_windx\_mean\_0.4 &                x &                     &            x \\
     mean\_windx\_min\_0.1 &                x &                     &            x \\
     mean\_windx\_min\_0.2 &                x &                     &            x \\
     mean\_windx\_min\_0.4 &                x &                   x &            x \\
             mean\_windy &                x &                   x &            x \\
     mean\_windy\_max\_0.1 &                x &                   x &            x \\
     mean\_windy\_max\_0.2 &                x &                     &            x \\
     mean\_windy\_max\_0.4 &                x &                   x &            x \\
    mean\_windy\_mean\_0.1 &                x &                     &            x \\
    mean\_windy\_mean\_0.2 &                x &                     &            x \\
    mean\_windy\_mean\_0.4 &                x &                     &            x \\
     mean\_windy\_min\_0.1 &                x &                     &            x \\
     mean\_windy\_min\_0.2 &                x &                   x &            x \\
     mean\_windy\_min\_0.4 &                x &                     &            x \\
              min\_iceaf &                  &                     &            x \\
      min\_iceaf\_max\_0.1 &                  &                     &            x \\
      min\_iceaf\_max\_0.2 &                  &                     &            x \\
      min\_iceaf\_max\_0.4 &                  &                     &            x \\
     min\_iceaf\_mean\_0.1 &                  &                     &            x \\
     min\_iceaf\_mean\_0.2 &                  &                     &            x \\
     min\_iceaf\_mean\_0.4 &                  &                     &            x \\
      min\_iceaf\_min\_0.1 &                  &                     &            x \\
      min\_iceaf\_min\_0.2 &                  &                     &            x \\
      min\_iceaf\_min\_0.4 &                  &                     &            x \\
          min\_magnitude &                x &                   x &            x \\
  min\_magnitude\_max\_0.1 &                x &                   x &            x \\
  min\_magnitude\_max\_0.2 &                x &                     &            x \\
  min\_magnitude\_max\_0.4 &                x &                     &            x \\
 min\_magnitude\_mean\_0.1 &                x &                     &            x \\
 min\_magnitude\_mean\_0.2 &                x &                     &            x \\
 min\_magnitude\_mean\_0.4 &                x &                     &            x \\
  min\_magnitude\_min\_0.1 &                x &                     &            x \\
  min\_magnitude\_min\_0.2 &                x &                     &            x \\
  min\_magnitude\_min\_0.4 &                x &                     &            x \\
           min\_pressure &                x &                     &            x \\
   min\_pressure\_max\_0.1 &                x &                     &            x \\
   min\_pressure\_max\_0.2 &                x &                     &            x \\
   min\_pressure\_max\_0.4 &                x &                     &            x \\
  min\_pressure\_mean\_0.1 &                x &                   x &            x \\
  min\_pressure\_mean\_0.2 &                x &                   x &            x \\
  min\_pressure\_mean\_0.4 &                x &                     &            x \\
   min\_pressure\_min\_0.1 &                x &                     &            x \\
   min\_pressure\_min\_0.2 &                x &                     &            x \\
   min\_pressure\_min\_0.4 &                x &                     &            x \\
              min\_windx &                x &                   x &            x \\
      min\_windx\_max\_0.1 &                x &                   x &            x \\
      min\_windx\_max\_0.2 &                x &                   x &            x \\
      min\_windx\_max\_0.4 &                x &                   x &            x \\
     min\_windx\_mean\_0.1 &                x &                     &            x \\
     min\_windx\_mean\_0.2 &                x &                     &            x \\
     min\_windx\_mean\_0.4 &                x &                     &            x \\
      min\_windx\_min\_0.1 &                x &                     &            x \\
      min\_windx\_min\_0.2 &                x &                   x &            x \\
      min\_windx\_min\_0.4 &                x &                     &            x \\
              min\_windy &                x &                   x &            x \\
      min\_windy\_max\_0.1 &                x &                   x &            x \\
      min\_windy\_max\_0.2 &                x &                   x &            x \\
      min\_windy\_max\_0.4 &                x &                   x &            x \\
     min\_windy\_mean\_0.1 &                x &                     &            x \\
     min\_windy\_mean\_0.2 &                x &                     &            x \\
     min\_windy\_mean\_0.4 &                x &                   x &            x \\
      min\_windy\_min\_0.1 &                x &                     &            x \\
      min\_windy\_min\_0.2 &                x &                     &            x \\
      min\_windy\_min\_0.4 &                x &                     &            x \\
\hline
%\bottomrule
%\end{tabular}
\end{longtable}

\bibliographystyle{elsarticle-num}
\bibliography{main}  %%% Uncomment this line and comment out the ``thebibliography'' section below to use the external .bib file (using bibtex) .

\end{document}